\documentclass[prd, twocolumn]{revtex4-2}
\usepackage{amsmath}
\usepackage{amssymb}
\usepackage{enumitem}
\usepackage{graphicx}

\allowdisplaybreaks

\begin{document}
%\title{Testing an inequality for relativistic local quantum measurements}
%\title{Towards an experimental test for a relativistic local quantum measurement inequality}
%\title{Experimental prospects for a relativistic local quantum measurement inequality}
\title{Realistic prospects for testing a relativistic local quantum measurement inequality}

\author{Riccardo Falcone}
\affiliation{Department of Physics, University of Sapienza, Piazzale Aldo Moro 5, 00185 Rome, Italy}

\author{Claudio Conti}
\affiliation{Department of Physics, University of Sapienza, Piazzale Aldo Moro 5, 00185 Rome, Italy}

\begin{abstract}
We investigate the experimental prospects for testing a relativistic local quantum measurement inequality that quantifies the trade-off between vacuum insensitivity and responsiveness to excitations for finite-size detectors. Building on the Reeh--Schlieder approximation for coherent states, we derive an explicit and practically applicable bound for arbitrary coherent states. To connect with realistic photodetection scenarios, we model the detection region as a square prism operating over a finite time window and consider a normally incident single-mode coherent state. Numerical results exhibit the expected qualitative behavior: suppressing dark counts necessarily tightens the achievable click probability.
\end{abstract}

\maketitle

\section{Introduction}

In our previous work \cite{falcone2025inequalityrelativisticlocalquantum}, we derived an inequality that quantifies the trade-off between vacuum insensitivity and responsiveness to excitations for finite-size detectors. Modeling a detector click by a local POVM element $\hat{E}_\text{click}$ supported in the finite region $\mathcal{O}_\text{det}$, and denoting the vacuum by $| \Omega \rangle$, locality implies that for any state $| \psi \rangle$, the click probability $P_\text{click} = \langle \psi | \hat{E}_\text{click} | \psi \rangle$ and the dark-count probability $P_\text{dark} = \langle \Omega | \hat{E}_\text{click} | \Omega \rangle$ are constrained by
\begin{equation}\label{upper_bound}
P_\text{click}  \leq \left( \mathcal{E}_\zeta  + \left\| \hat{A}_\zeta  \right\| \sqrt{P_\text{dark}} \right)^2.
\end{equation}
Here, $\hat{A}_\zeta $ is a one-parameter family of operators localized in the causal complement $\mathcal{O}_\text{det}'$ that approximate $| \psi \rangle$ by acting on the vacuum $| \Omega \rangle$, with error
\begin{equation}\label{cyclicity}
\mathcal{E}_\zeta  = \left\| | \psi \rangle - \hat{A}_\zeta  | \Omega \rangle  \right\|
\end{equation}
tending to zero as $\zeta \to 0$. The existence of such operators $\hat{A}_\zeta$ is guaranteed by the Reeh--Schlieder theorem \cite{Reeh1961, haag1992local}. Different choices of $\hat{A}_\zeta$ yield different bounds in \eqref{upper_bound}; the optimal inequality is obtained by minimizing the right-hand side over all admissible families $\{ \hat{A}_\zeta \}$ and all values of $\zeta$.

In this paper, we examine the practical prospects for testing the inequality \eqref{upper_bound} in realistic experimental settings. Unlike Ref.~\cite{falcone2025inequalityrelativisticlocalquantum}, which considered a toy setup with co-centered hyperspherical detector and coherent state, we instead model a prismatic detector operated over a fixed time interval and take $| \psi \rangle$ to be a single-mode coherent state. These choices are intended to approximate practical photodetectors and laser beams in laboratory settings.

To obtain an explicit form of Eq.~\eqref{upper_bound}, we build on Ref.~\cite{falcone2025reehschliederapproximation}, which constructs a one-parameter family of operators $\hat{A}_\zeta$ localized in the causal complement of an arbitrary region and capable of approximating any coherent state by acting on $| \Omega \rangle$. We model the detection region $\mathcal{O}_\text{det}$ as a hyperrectangle and take $| \psi \rangle$ to be a single-mode coherent state. For each configuration (specified by the detector size, coherent-state amplitude, frequency, and coherence volume) we derive a suboptimal bound from \eqref{upper_bound} by minimizing with respect to $\zeta$. 

The paper is organized as follows. In Sec.~\ref{ReehSchlieder_approximation_for_coherent_states} we review the construction from Ref.~\cite{falcone2025reehschliederapproximation} of Reeh--Schlieder operators $\hat{A}_\zeta$ approximating any arbitrary coherent state $| f \rangle$. In Sec.~\ref{Inequality_for_general_coherent_states}, we use this construction to obtain an explicit form of \eqref{upper_bound} when $| \psi \rangle = | f \rangle$. In Sec.~\ref{Reformulating_the_results_in_terms_of_coherentstate_amplitudes}, we rewrite the bound in terms of the coherent-state amplitude rather than the smearing function $f$, making contact with experimental scenarios. Section \ref{right_square_prism_detector_and_a_normally_incident_single_mode_coherent_state} specializes to a concrete setting, taking $\mathcal{O}_\text{det}$ to be a hyperrectangle and $| \psi \rangle$ a single-mode coherent state, and derives the corresponding bound. %Numerical results are presented in Sec.~\ref{Numerical_results}.
Conclusions are drawn in Sec.~\ref{Conclusions}. %Details of the numerical computations used in Sec.~\ref{right_square_prism_detector_and_a_normally_incident_single_mode_coherent_state} are provided in Appendix \ref{Specifics_of_the_numerical_computation}.

\section{Reeh--Schlieder approximation for coherent states}\label{ReehSchlieder_approximation_for_coherent_states}

In Ref.~\cite{falcone2025reehschliederapproximation}, we considered an arbitrary coherent state $| f \rangle$ and a spacetime region $\mathcal{U}$, and constructed a one-parameter family of operators $\hat{A}_\zeta$, localized in the causal complement of $\mathcal{U}$, that approximate $| f \rangle$ by acting on the vacuum $| \Omega \rangle$, with the approximation error tending to zero as $\zeta \to 0$. In this section, we briefly review that construction, tailoring it to the present setting in which the detector region $\mathcal{O}_\text{det}$ plays the role of $\mathcal{U}$. 

The coherent state $| f \rangle $ is defined as $| f \rangle = \hat{W}(f) | \Omega \rangle$, where
\begin{equation}\label{W}
\hat{W}(f) = \exp \left[i \hat{\phi}(f) \right]
\end{equation}
is the Weyl operator associated with the test function $f$, and $\hat{\phi}(f) $ is the field operator smeared with $f$, i.e.,
\begin{equation}
\hat{\phi}(f) = \int_{\mathbb{R}^4} d^4 x f(x) \hat{\phi}(x).
\end{equation}
Here, for simplicity, we allow ourselves the notational shortcut of using $\hat{\phi}(f)$ for the smeared field operator and $\hat{\phi}(x)$ for the pointwise field operator.

Let $\mathcal{O}_\text{det}'' = (\mathcal{O}_\text{det}')'$ denote the causal completion of $\mathcal{O}_\text{det}$. We enlarge this region to an open set $\mathcal{O}_1  \supset \mathcal{O}_\text{det}''$ so that the union $\mathcal{O}_1 \cup \mathcal{O}_\text{det}'$ contains a Cauchy surface for the full spacetime. We choose a region $\mathcal{O}_1$ and a coordinate system such that this Cauchy surface is identified by $x^0 = 0$. By the time-slice property of the field \cite{Dimock1980}, the presence of this Cauchy surface ensures the existence of a test function $f_0$ supported entirely within $\mathcal{O}_1 \cup \mathcal{O}_\text{det}'$ such that $\hat{W}(f) = \hat{W}(f_0)$.

We further enlarge to a region $\mathcal{O}_2 \supset \mathcal{O}_1$ and consider a smooth bump $\chi$ with $0\leq\chi\leq 1$, $\chi |_{\mathcal{O}_1} = 1$, $\chi|_{\mathcal{O}_2}\neq 0$ and $\chi(x)=0$ for all $x\notin \mathcal{O}_2$, so that test functions are smoothly localized to $\mathcal{O}_1$ using the collar region $\mathcal{O}_2 \setminus \mathcal{O}_1$ as a transition zone. The complementary cutoff $\chi' = 1 - \chi$ provides the decomposition $f_0 = \chi f_0+ \chi' f_0$, with $\chi f_0$ supported in $\mathcal{O}_2 \cap (\mathcal{O}_1 \cup \mathcal{O}_\text{det}')$ and $\chi' f_0$ supported in $\mathcal{O}_\text{det}' \setminus \mathcal{O}_1$. We choose spacetime coordinates $(x^0, x^1, x^2, x^3)$ so that the left wedge $\mathcal{W}_\text{L} = \{ x : x^1<-|x^0| \} $ contains $\mathcal{O}_2$.

Building on Ref.~\cite{falcone2025reehschliederapproximation}, we obtain the family of operators
%\begin{align}\label{A_zeta_f}
%\hat{A}_\zeta(f) = & \exp\!\left\lbrace i \Im\!\left[ W_2\!\left(\chi' f_0,\chi f_0\right) \right] \right\rbrace  \hat{W}(\chi' f_0) \nonumber \\
%& \times \int_\mathbb{R} d\eta \, G_\zeta(\eta - i \pi) \, \hat{W}\!\left[\chi f_0\circ J \circ \Lambda_1(-\eta)\right],
%\end{align}
%which are localized in $\mathcal{O}_\text{det}'$ and, acting on the vacuum $| \Omega \rangle$, approximate the target state $| f \rangle$ with error
\begin{widetext}
\begin{equation}\label{A_zeta_f}
\hat{A}_\zeta(f) = \exp\!\left\lbrace i \Im\!\left[ W_2\!\left(\chi' f_0,\chi f_0\right) \right] \right\rbrace  \hat{W}(\chi' f_0)\int_\mathbb{R} d\eta \, G_\zeta(\eta - i \pi) \, \hat{W}\!\left[\chi f_0\circ J \circ \Lambda_1(-\eta)\right],
\end{equation}
which are localized in $\mathcal{O}_\text{det}'$ and, acting on the vacuum $| \Omega \rangle$, approximate the target state $| f \rangle$ with error
\begin{equation}\label{epsilon_zeta_F_2}
\mathcal{E}_\zeta(f) = \sqrt{1 - \int_\mathbb{R} d\eta  [ 2 G_\zeta(\eta) -  G_{2 \zeta}(\eta)]  \exp \left\lbrace W_2 \! \left[ \chi f_0,  \chi f_0\circ \Lambda_1(\eta) \right] - W_2 (\chi f_0,\chi f_0) \right\rbrace }.
\end{equation} 
\end{widetext}
Here,
\begin{equation}
G_\zeta(\eta) = \frac{1}{\sqrt{2 \pi \zeta}} \exp \left(- \frac{\eta^2}{ 2 \zeta} \right)
\end{equation}
is a Gaussian function with variance $\zeta$,
\begin{equation}
W_2(f_1,f_2) = \left\langle \Omega \middle| \hat{\phi}(f_1) \hat{\phi}(f_2) \middle| \Omega \right\rangle
\end{equation}
is the two-point Wightman function smeared with $f_1$ and $f_2$, $J$ is the spacetime reflection about the $x^0$ and $x^1$ directions,
\begin{equation}
J (x^0, x^1, x^2, x^3) = (-x^0, -x^1, x^2, x^3),
\end{equation}
and $\Lambda_1(\eta)$ denotes a Lorentz boost along $x^1$,
\begin{align}\label{Lambda_1_eta}
& \Lambda_1(\eta)\!\left(x^0, x^1, x^2, x^3\right) = \left[\cosh (\eta) x^0 + \sinh (\eta) x^1,  \right. \nonumber\\ 
& \left. \cosh (\eta) x^1 + \sinh (\eta) x^0, x^2, x^3\right].
\end{align}

\section{Inequality for coherent states}\label{Inequality_for_general_coherent_states}

In Ref.~\cite{falcone2025inequalityrelativisticlocalquantum}, we gave an explicit form of the inequality \eqref{upper_bound} in a specific setting: a coherent state $| f \rangle = \hat{W}(f) | \Omega \rangle$ with $\hat{W}(f)$ localized in the causal completion $\mathcal{O}_\text{det}''$. In the present work, we drop this locality assumption on $f$. Allowing arbitrary support, we use the results summarized in Sec.~\ref{ReehSchlieder_approximation_for_coherent_states} to derive an explicit form of the inequality \eqref{upper_bound} for this more general class of states.

By using the triangle inequality on \eqref{A_zeta_f} and the fact that unitary operators have norm $1$, we derive an upper bound on the operator norm of $ \hat{A}_\zeta(f) $, that reads
\begin{equation}\label{W_prime_zeta_norm_upper_bound_Gaussian}
\left\| \hat{A}_\zeta(f) \right\| \leq \exp\! \left( \frac{\pi^2}{2\zeta}\right).
\end{equation}
Plugging Eqs.~\eqref{epsilon_zeta_F_2} and \eqref{W_prime_zeta_norm_upper_bound_Gaussian} into the inequality \eqref{upper_bound}, with the identifications $P_\text{click}   \mapsto P_\text{click} (f)$, $\mathcal{E}_\zeta  \mapsto \mathcal{E}_\zeta (f) $ and $\hat{A}_\zeta  \mapsto \hat{A}_\zeta(f)$, yields
\begin{equation}\label{upper_bound_f_O_d_1}
P_\text{click} (f)  \leq \left[ \mathcal{E}_\zeta (f) + \exp\! \left( \frac{\pi^2}{2\zeta}\right) \sqrt{P_\text{dark}} \right]^2.
\end{equation}

The inequality holds for every $\zeta > 0$. To obtain the tightest bound achievable within the chosen family of operators $\{ \hat{A}_\zeta \}$, we minimize the right-hand side of \eqref{upper_bound_f_O_d_1} with respect to $\zeta>0$:
\begin{equation}\label{upper_bound_f_O_d_2}
P_\text{click} (f)  \leq \min_{\zeta>0} \left[ \mathcal{E}_\zeta (f) + \exp\! \left( \frac{\pi^2}{2\zeta}\right) \sqrt{P_\text{dark}} \right]^2.
\end{equation}

The form of Eq.~\eqref{upper_bound_f_O_d_2} is identical to what we found in Ref.~\cite{falcone2025inequalityrelativisticlocalquantum} for the case in which $\hat{W}(f)$ is localized within $\mathcal{O}_\text{det}''$. However, the expression for $\mathcal{E}_\zeta (f)$ now differs, as it is now given by Eq.~\eqref{epsilon_zeta_F_2}.

In this setting, the bound depends on the specific choices of the extended regions $\mathcal{O}_1$ and $\mathcal{O}_2$ as well as the cutoff function $\chi$. Ideally, these should be chosen to make the bound as tight as possible, although identifying the optimal configuration is not always straightforward. A helpful rule of thumb is this: shrinking the collar $\mathcal{O}_2 \setminus \mathcal{O}_\text{det}$ tends to reduce the contribution from $f_0$, since the factor $\chi$ suppresses the field over a larger portion of spacetime. If $\mathcal{O}_2$ is taken too large relative to $\mathcal{O}_\text{det}$, the dependence on the detector finite size may be lost, effectively turning the measurement into a global one and making the inequality practically unusable.

At the same time, making the collar too thin steepens the cutoff. In the extreme limit $\mathcal{O}_2 \to \mathcal{O}_\text{det}$, the bump turns into the characteristic function of $\mathcal{O}_\text{det}$, and $ \chi f_0$ loses smoothness at the boundary. That lack of smoothness can blow up the Wightman two-point term $W_2[\chi f_0,\chi f_0\circ \Lambda_1(\eta)]$ inside the error $\mathcal{E}_\zeta(f)$ and thereby sabotage the bound in Eq.~\eqref{upper_bound_f_O_d_2}.

In practice, therefore, one has to find a trade-off: the collar must be narrow enough to suppress unwanted support, yet wide enough to keep $ \chi f_0$ smooth so that the two-point function remains finite. One may reasonably assume that the optimal width of the collar $\mathcal{O}_2 \setminus \mathcal{O}_\text{det}$ should be neither significantly larger nor smaller than the size of the detector.

The coordinates $x$ can be chosen arbitrarily so long as $\mathcal{O}_2$ lies entirely within the left wedge $\mathcal{W}_\text{L}$. To sharpen the bound in \eqref{upper_bound_f_O_d_2}, it is advantageous to choose coordinates that minimize its right-hand side. A practical choice is to place $\mathcal{O}_2$ as close to the origin as possible; this reduces the effect of the boost $\Lambda_1(\eta)$ in \eqref{epsilon_zeta_F_2}, thereby decreasing $\mathcal{E}_\zeta (f)$.

%\section{Reformulating the results in terms of coherent-state amplitudes}\label{Reformulating_the_results_in_terms_of_coherentstate_amplitudes}
\section{Expressing the results via coherent-state amplitudes}\label{Reformulating_the_results_in_terms_of_coherentstate_amplitudes}

Coherent states are typically expressed using the multi-mode displacement operator, defined as
\begin{equation}\label{D_alpha}
\hat{D}(\alpha) = \exp \left\lbrace \int_{\mathbb{R}^3} d^3\mathbf{k} \left[ \alpha(\mathbf{k}) \hat{a}^\dagger(\mathbf{k}) - \alpha^*(\mathbf{k}) \hat{a}(\mathbf{k}) \right] \right\rbrace,
\end{equation}
where $\alpha(\mathbf{k}) $ is the coherent-state amplitude and $\hat{a}(\mathbf{k}) $ denotes the annihilation operators associated to the momentum $\mathbf{k}$. Using this definition, coherent states are written as $| \alpha \rangle = \hat{D}(\alpha) | \Omega \rangle$. This representation is more familiar across various areas of physics, including quantum optics, condensed matter, and continuous-variable quantum information. Expressing coherent states in terms of $\alpha$ facilitates both their interpretation and their relevance to experimental scenarios.

In this section, we recast the results of Sec.~\ref{Inequality_for_general_coherent_states} in this representation, rewriting \eqref{upper_bound_f_O_d_2} directly in terms of the amplitude $\alpha$. This allows us to connect more transparently with experimental settings.

\subsection{Coherent states as displaced vacuum states}\label{Coherent_states_as_displaced_vacuum_states}

In this subsection, we clarify the connection between coherent states in Algebraic Quantum Field Theory (AQFT), defined as $| f \rangle = \hat{W}(f) | \Omega \rangle$, and the notion of coherent states $| \alpha \rangle = \hat{D}(\alpha) | \Omega \rangle$ constructed using the multi-mode displacement operator \eqref{D_alpha}.

Let us start from the expression of the Weyl operator given in Eq.~\eqref{W}. Expanding the field operator $\hat{\phi}$ in terms of Minkowski modes and their corresponding annihilation and creation operators yields
\begin{equation}\label{phi_u_a}
\hat{\phi}(x) = \int_{\mathbb{R}^3} d^3 \mathbf{k} \left[ u(\mathbf{k}, x) \hat{a}(\mathbf{k}) + u^*(\mathbf{k},x) \hat{a}^\dagger(\mathbf{k})  \right].
\end{equation}
The free modes $u $ are defined as
\begin{equation}\label{u}
u(\mathbf{k}, x) =  \frac{e^{-i\omega(\mathbf{k})x^0 + i\mathbf{k} \cdot \mathbf{x}}}{\sqrt{(2\pi)^3 2 \omega(\mathbf{k})}},
\end{equation}
where $\omega(\mathbf{k}) = \sqrt{m^2 + |\mathbf{k}|^2}$ denotes the energy associated with momentum $\mathbf{k}$ and $m$ is the mass of the field.

By substituting Eq.~\eqref{phi_u_a} into Eq.~\eqref{W}, we find that the identity $\hat{W}(f) = \hat{D}(\alpha)$ holds when
\begin{equation}\label{alpha_f}
\alpha(\mathbf{k}) = i \int_{\mathbb{R}^4} d^4 x f(x)  u^*(\mathbf{k}, x).
\end{equation}
This establishes a direct correspondence between the two formulations of coherent states: $\hat{W}(f) | \Omega \rangle = \hat{D}(\alpha) | \Omega \rangle$.

\subsection{Initial-data smearing function $f_0$ in terms of the coherent-state amplitude $\alpha$}

In Sec.~\ref{ReehSchlieder_approximation_for_coherent_states} we introduced a representation of coherent states in terms of the initial-data smearing function $f_0$. Then, in Sec.~\ref{Coherent_states_as_displaced_vacuum_states}, we established the connection between the Weyl-operator-based representation $\hat{W}(f)$ and the more familiar one involving the displacement operator $\hat{D}(\alpha)$. In this subsection, we directly connect the latter representation to the one based on the functions $f_0$. Concretely, we derive an explicit expression for $f_0$ in terms of the amplitude $\alpha$. Combined with Eq.~\eqref{epsilon_zeta_F_2}, this leads to a concrete expression of the inequality \eqref{upper_bound_f_O_d_2}, making it directly applicable to experimental analysis.

We chose the coordinate system such that the hypersurface $x^0 = 0$ is contained within $\mathcal{O}_1 \cup \mathcal{O}_\text{det}'$. Under this assumption, for any extension $\mathcal{O}_1 \supset \mathcal{O}_\text{det}$ and sufficiently small $\varepsilon > 0$, the thin spacetime neighborhood $\Sigma_\varepsilon$, defined by $|x^0| < \varepsilon$, lies entirely within $\mathcal{O}_1 \cup \mathcal{O}_\text{det}'$. %As a result, the initial-data smearing function $f_0$ can be chosen to have support entirely contained within $\Sigma_\varepsilon$.

Let $| \alpha \rangle = \hat{D}(\alpha) | \Omega \rangle$ denote the coherent state with amplitude $\alpha$. Pick a real, smooth bump function $\delta_\varepsilon$ on $\mathbb{R}$ such that $\operatorname{supp}(\delta_\varepsilon) \subseteq (-\varepsilon, \varepsilon)$ and $\int_{\mathbb{R}} dx^0 \delta_\varepsilon(x^0) = 1$. For any such choice, the real-valued function
\begin{align}\label{f_0_alpha_varepsilon}
f_0(x) = & \frac{\partial}{\partial x^0} \left\lbrace \delta_\varepsilon(x^0) \int_{\mathbb{R}^3} d^3\mathbf{k} \, 2 \Re [\alpha(\mathbf{k}) u(\mathbf{k}, x)] \right\rbrace \nonumber \\
& + \delta_\varepsilon(x^0) \int_{\mathbb{R}^3} d^3\mathbf{k} \, 2 \Im [\alpha(\mathbf{k}) \omega(\mathbf{k}) u(\mathbf{k}, x)]
\end{align}
satisfies all the necessary conditions to serve as an initial-data smearing function corresponding to the coherent state $| \alpha \rangle$. Indeed, by integrating by parts and using the explicit form of the mode functions $u $ [Eq.~\eqref{u}], along with the normalization$\int_{\mathbb{R}} dx^0 \delta_\varepsilon(x^0) = 1$, one can show that $f_0$ satisfies Eq.~\eqref{alpha_f} with $f_0$ taking the role of $f$. This confirms the equivalence $ \hat{D}(\alpha) | \Omega \rangle = \hat{W}(f_0) | \Omega \rangle$. Also, since $\operatorname{supp}(\delta_\varepsilon) \subseteq (-\varepsilon, \varepsilon)$, it follows that $f_0$ is supported within the neighborhood $\Sigma_\varepsilon$, which is contained in $\mathcal{O}_1 \cup \mathcal{O}_\text{det}'$. As a result, Eq.~\eqref{f_0_alpha_varepsilon} explicitly establishes the connection between the representation of the coherent state in terms of the amplitude $\alpha$ and the one based on the initial-data smearing function $f_0$.

In the limit of $\varepsilon \to 0$, any choice of $f_0$ constructed as in Eq.~\eqref{f_0_alpha_varepsilon} converges to
\begin{align}\label{f_0_alpha}
f_0(x) = & \frac{\partial}{\partial x^0} \left\lbrace \delta(x^0) \int_{\mathbb{R}^3} d^3\mathbf{k} \, 2 \Re [\alpha(\mathbf{k}) u(\mathbf{k}, x)] \right\rbrace \nonumber \\
& + \delta(x^0) \int_{\mathbb{R}^3} d^3\mathbf{k} \, 2 \Im [\alpha(\mathbf{k}) \omega(\mathbf{k}) u(\mathbf{k}, x)],
\end{align}
where $\delta$ denotes the Dirac delta distribution. It is important to note that this expression for $f_0$ is no longer a smooth function in the conventional sense, but rather a distribution supported on the hypersurface $x^0=0$. As such, it does not strictly belong to the space of admissible smearing functions in AQFT, where localization must occur over open spacetime regions. To remain mathematically rigorous within AQFT, one should work with the regularized expression in Eq.~\eqref{f_0_alpha_varepsilon} for all computations involving $f_0$. The outcome will ultimately be independent of the specific regularization $\delta_\varepsilon$, since the coherent state $\hat{W}(f_0) | \Omega \rangle$ itself is independent of this choice. Nevertheless, in what follows, we will assume that all operations involving $f_0$ commute with the limit $\varepsilon \to 0$. Therefore, taking this limit now rather than later---when $\varepsilon$ would disappear from the expressions---is entirely justified. This allows us to adopt a heuristic approach based on the simplified form in Eq.~\eqref{f_0_alpha}. As long as the limit $\varepsilon \to 0$ continues to commute with the operations we perform, this treatment remains valid and consistent.

Also, when taking the limit $\varepsilon \to 0$, we may simultaneously consider the limit $\mathcal{O}_1 \to \mathcal{O}_\text{det}''$, since, for each smaller $\varepsilon$, the support of $f_0$ remains contained in $\mathcal{O}_1 \cup \mathcal{O}_\text{det}'$ for any sufficiently thin collar $\mathcal{O}_1 \cap \mathcal{O}_\text{det}''$, so in the limit the collar can be made arbitrarily narrow. Heuristically, thinning the collar reduces the contribution of $f_0$ to the error $\mathcal{E}_\zeta(f)$, thereby tightening the bound \eqref{upper_bound_f_O_d_2}.

\subsection{Boosted Wightman two-point function $W_2[\chi f_0,\chi f_0\circ \Lambda_1(\eta)]$ in terms of the coherent-state amplitude $\alpha$}

Equation \eqref{f_0_alpha} provides an explicit expression for the initial-data smearing function $f_0$ in terms of the coherent-state amplitude $\alpha$. This allows us to rewrite the boosted Wightman two-point function $W_2[\chi f_0,\chi f_0\circ \Lambda_1(\eta)] = \langle \Omega | \hat{\phi}(\chi f_0) \hat{\phi}[\chi f_0\circ \Lambda_1(\eta)] | \Omega \rangle$ directly in terms of $\alpha$. To obtain this result, we start from the explicit expression of $W_2(f_1,f_2)$ in terms of the arbitrary smearing functions $f_1$ and $f_2$:
\begin{align}\label{W_2_f_1_f_2}
& W_2(f_1,f_2)\nonumber \\
= & \int_{\mathbb{R}^3} d^3 \mathbf{k} \int_{\mathbb{R}^4} d^4 x \int_{\mathbb{R}^4} d^4 y f_1(x) f_2(y) u(\mathbf{k}, x) u^*(\mathbf{k}, y),
\end{align}
which follows from Eq.~\eqref{phi_u_a}, the canonical commutation relation $[\hat{a}(\mathbf{k}), \hat{a}^\dagger(\mathbf{k}')] = \delta^3(\mathbf{k}-\mathbf{k}')$ and the vacuum condition $\hat{a}(\mathbf{k}) | \Omega \rangle = 0$. In our case, we take $f_1 = \chi f_0$ and $f_2 = \chi f_0\circ \Lambda_1(\eta)$.

To proceed, we apply coordinate transformations and integrate by parts to shift the action of the Lorentz boost $ \Lambda_1(\eta)$ and the time derivatives $\partial/\partial x^0$ and $\partial/\partial y^0$, onto the mode functions $ u(\mathbf{k}, x)$ and $ u(\mathbf{k}, y)$. This leads to the following expression:
\begin{widetext}
\begin{align}\label{W_2_f_d_Lambda_eta_f_d}
& W_2[\chi f_0,\chi f_0\circ \Lambda_1(\eta)]  = \int_{\mathbb{R}^3} d^3 \mathbf{k} \int_{\mathbb{R}^3} d^3 \mathbf{p} \int_{\mathbb{R}^3} d^3 \mathbf{q} \int_{\mathbb{R}^3} d^3 \mathbf{x} \int_{\mathbb{R}^3} d^3 \mathbf{y} \chi(x) \chi(y) \left\lbrace 2 \Im [\alpha(\mathbf{p}) \omega(\mathbf{p}) u(\mathbf{p}, x)] \right.  \nonumber \\
 & \left. - 2 \Re [\alpha(\mathbf{p}) u(\mathbf{p}, x)] \frac{\partial}{\partial x^0} \right\rbrace \left. \left\lbrace 2 \Im [\alpha(\mathbf{q}) \omega(\mathbf{q}) u(\mathbf{q}, y)]  - 2 \Re [\alpha(\mathbf{q}) u(\mathbf{q}, y)] \frac{\partial}{\partial y^0} \right\rbrace  u(\mathbf{k}, x) u^*\!\left[\mathbf{k}, \Lambda_1^{-1}(\eta)(y)\right] \right|_{x^0=y^0=0}.
\end{align}
Here, for simplicity, we have assumed that the bump function $\chi$ is chosen such that $\partial\chi/\partial x^0 |_{x^0 = 0} = 0$.

A more explicit form of Eq.~\eqref{W_2_f_d_Lambda_eta_f_d} can be obtained by using Eqs.~\eqref{Lambda_1_eta} and \eqref{u}, resulting in
\begin{align}\label{W_2_f_d_Lambda_eta_f_d_2}
W_2[\chi f_0,\chi f_0\circ \Lambda_1(\eta)]
% = & \frac{1}{(2\pi)^6} \int_{\mathbb{R}^3}  \frac{d^3 \mathbf{k}}{ \omega(\mathbf{k})} \int_{\mathbb{R}^3} \frac{d^3 \mathbf{p}}{\sqrt{\omega(\mathbf{p})}} \int_{\mathbb{R}^3} d^3 \mathbf{x} \chi(0,\mathbf{x}) \left\lbrace  \omega(\mathbf{p}) \Im \!\left[\alpha(\mathbf{p}) e^{i \mathbf{p} \cdot \mathbf{x}}\right]  + i \omega(\mathbf{k}) \Re \!\left[\alpha(\mathbf{p}) e^{i \mathbf{p} \cdot \mathbf{x}}\right]  \right\rbrace e^{i \mathbf{k} \cdot \mathbf{x}} \nonumber \\
% & \times  \int_{\mathbb{R}^3} \frac{d^3 \mathbf{q}}{\sqrt{\omega(\mathbf{q})}} \int_{\mathbb{R}^3} d^3 \mathbf{y} \chi(0,\mathbf{y})\left\lbrace \omega(\mathbf{q}) \Im \!\left[\alpha(\mathbf{q}) e^{i \mathbf{q} \cdot \mathbf{y}}\right]  - \Re \!\left[\alpha(\mathbf{q}) e^{i \mathbf{q} \cdot \mathbf{y}}\right] \frac{\partial}{\partial y^0} \right\rbrace\nonumber \\
% & \times  \left.  \exp \left\lbrace i\omega(\mathbf{k})\left[\cosh (\eta) y^0 - \sinh (\eta) y^1\right] - i k^1 \left[\cosh (\eta) y^1 - \sinh (\eta) y^0\right] - i\mathbf{k}_\perp \cdot \mathbf{y}_\perp \right\rbrace  \right|_{y^0=0}\nonumber \\
 = & \frac{1}{(2\pi)^6} \int_{\mathbb{R}^3}  \frac{d^3 \mathbf{k}}{ \omega(\mathbf{k})} \int_{\mathbb{R}^3} \frac{d^3 \mathbf{p}}{\sqrt{\omega(\mathbf{p})}} \int_{\mathbb{R}^3} d^3 \mathbf{x} \chi(0,\mathbf{x}) \left\lbrace \omega(\mathbf{p}) \Im \!\left[\alpha(\mathbf{p}) e^{i \mathbf{p} \cdot \mathbf{x}}\right]  + i \omega(\mathbf{k}) \Re \!\left[\alpha(\mathbf{p}) e^{i \mathbf{p} \cdot \mathbf{x}}\right]  \right\rbrace e^{i \mathbf{k} \cdot \mathbf{x}} \nonumber \\
& \times  \int_{\mathbb{R}^3} \frac{d^3 \mathbf{q}}{\sqrt{\omega(\mathbf{q})}} \int_{\mathbb{R}^3} d^3 \mathbf{y} \chi(0,\mathbf{y})\left\lbrace \omega(\mathbf{q}) \Im \!\left[\alpha(\mathbf{q}) e^{i \mathbf{q} \cdot \mathbf{y}}\right]  - i \left[ \cosh (\eta) \omega(\mathbf{k}) + \sinh (\eta) k^1 \right] \Re \!\left[\alpha(\mathbf{q}) e^{i \mathbf{q} \cdot \mathbf{y}}\right]  \right\rbrace\nonumber \\
& \times   \exp \! \left\lbrace -i \left[ \sinh (\eta) \omega(\mathbf{k}) + \cosh (\eta) k^1 \right] y^1 - i\mathbf{k}_\perp \cdot \mathbf{y}_\perp \right\rbrace,
\end{align}
\end{widetext}
where $\mathbf{k}_\perp = (k^2, k^3)$ and $\mathbf{y}_\perp = (y^2, y^3)$ denote the transverse spatial components of the momentum $\mathbf{k}$ and the Minkowski coordinate $y$, respectively. This explicit expression can be used in combination with Eq.~\eqref{epsilon_zeta_F_2} and $W_2(\chi f_0, \chi f_0) = W_2[\chi f_0,\Lambda_1(0) \chi f_0]$ to compute the right-hand side of inequality \eqref{upper_bound_f_O_d_2} in terms of $\alpha$.

\section{Square prism detector and normally incident single-mode coherent state}\label{right_square_prism_detector_and_a_normally_incident_single_mode_coherent_state}

To compute the inequality \eqref{upper_bound_f_O_d_2} explicitly, we must specify the detector region $\mathcal{O}_\text{det}$ and the smearing function $f$ defining the coherent state. Naturally, the size and the shape of the detector, as well as the coherent state profile, depend on the specifics of the experimental setup. The coordinate system $x$ can be chosen freely, as long as the support of $\chi$ lies entirely within the left wedge $\mathcal{W}_\text{L} = \{ x : |x^0|<-x^1 \}$. The choice of bump function $\chi$ is, in principle, arbitrary. However, to tighten the bound in Eq.~\eqref{upper_bound_f_O_d_2}, it is preferable to choose a form that optimizes the estimate. As discussed qualitatively in Sec.~\ref{Inequality_for_general_coherent_states}, this is expected to happen when the scale of $\chi$ beyond the detector boundaries is comparable to the size of the detector.

In this section, we examine a concrete setup in which the detector takes the shape of a rectangular prism with square base and dimensions $l \times L \times L$, where $l$ represents its thickness. The detector is assumed to operate at a fixed interval of time $\tau$. The corresponding spacetime region $\mathcal{O}_\text{det}$ is defined as the hyperrectangle with dimensions $\tau \times l \times L \times L$.

The field is prepared in a single-mode coherent state, specified by an amplitude $\alpha_0$, a momentum $\mathbf{k}_0$, a frequency $\omega_0 = \omega(\mathbf{k}_0)$ and coherence volume $V_\text{coh}$. We assume the mode is normally incident on the detector, meaning it propagates perpendicularly to the square faces of the prism. To avoid the nonrelativistic regime, where the implications of the Reeh--Schlieder theorem are suppressed \cite{FALCONE2024100095}, we focus exclusively on the massless case, $m=0$. Under this hypothesis, and neglecting polarization, the coherent state can represent a monochromatic electromagnetic wave that is locally planar, such as a Gaussian beam with a waist much larger than the detector itself. This configuration is, in principle, experimentally realizable within the context of quantum optics.

\begin{figure}[h]
\includegraphics[width=\columnwidth]{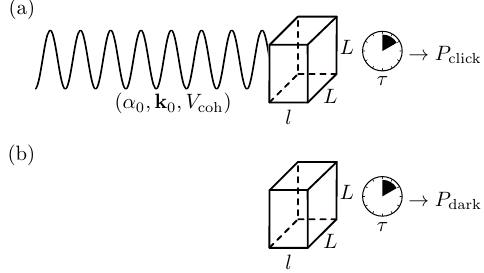}
\caption{In the proposed experiment, the same detector is used to measure the click probability $P_\text{click}$ of a chosen non-vacuum state $| \psi \rangle$ (panel a) and the dark-count probability $P_\text{dark}$ in the vacuum (panel b). The two outcomes are then compared with the theoretical bound relating these quantities [Eq.~\eqref{upper_bound_f_O_d_2}]. The detector region is modeled as a square prism of dimensions $l \times L \times L$ operating over a finite time window $\tau$, and the input state is taken to be a normally incident single-mode coherent state characterized by amplitude $\alpha_0$, momentum $\mathbf{k}_0$ and coherence volume $V_\text{coh}$.
}\label{Setup}
\end{figure}

A schematic of the proposed experiment is shown in Fig.~\ref{Setup}. The same square-prism detector is used to measure the click probability both in the presence of the coherent state and in the vacuum. The resulting values of $P_\text{click}$ and $P_\text{dark}$ are then expected to satisfy the constraint imposed by \eqref{upper_bound_f_O_d_2}.

\subsection{Construction of $\mathcal{O}_1$ and $\chi$, and fixing the coordinate system}

To construct the region $\mathcal{O}_1$, we extend the detector volume into a rectangular prism of dimensions $ (l + \tau) \times (L + \tau) \times (L + \tau)$ centered at the same point as $\mathcal{O}_\text{det}$. We denote this three-dimensional region by $\mathcal{V}_\text{det}$. Unlike the original detector volume of size $l \times L \times L$, $\mathcal{V}_\text{det}$ is enlarged by the measurement duration $\tau$. The region $\mathcal{O}_1$ is then defined as the causal completion of $\mathcal{V}_\text{det}$. By construction, $\mathcal{O}_1$ contains $\mathcal{O}_\text{det}''$ \footnote{We do not use $\mathcal{O}_\text{det}''$ itself as the enlarged region $\mathcal{O}_1$ purely for practical reasons, since integrating over a region with rectangular boundaries is simpler than over a three-dimensional slice of $\mathcal{O}_\text{det}''$.}. The support of the bump function $\chi$, denoted $\mathcal{O}_2$, is defined to contain $\mathcal{O}_1$. Its shape is chosen to match the diamond geometry of $\mathcal{O}_1$, but with dimensions enlarged by the margins $\delta l$ and $\delta L$. Accordingly, $\mathcal{O}_2$ is equivalent to the causal completion of a rectangle with dimensions $ (l + \tau + \delta l) \times (L + \tau + \delta L) \times (L + \tau + \delta L)$ sharing the same center as $\mathcal{O}_\text{det}$. 

The coordinate system $x$ is chosen such that the detector region $\mathcal{O}_\text{det}$ corresponds to the hyperrectangle $(-\tau/2, \tau/2) \times (-l - \tau/2 - \delta l/2, - \tau/2 - \delta l/2) \times (-L/2, L/2) \times (-L/2, L/2)$. The volume $\mathcal{V}_\text{det}$ (i.e., the three-dimensional cross-section of $\mathcal{O}_1$ at $x^0 = 0$) is given by the intervals $(-l - \tau - \delta l/2, - \delta l/2) \times (-L/2 - \tau/2, L/2 + \tau/2) \times (-L/2 - \tau/2, L/2 + \tau/2)$. Similarly, the support of the spatial function $ \chi |_{x^0=0}$ (i.e., the three-dimensional cross-section of $\mathcal{O}_2$ at $x^0 = 0$) is  $(-l  - \tau - \delta l, 0) \times (-L/2 - \tau/2 -\delta L/2, L/2 + \tau/2 +\delta L/2) \times (-L/2 - \tau/2 -\delta L/2, L/2 + \tau/2 +\delta L/2)$. With this coordinate choice, the momentum $\mathbf{k}_0$ is oriented along the $x^1$-axis, i.e., $\mathbf{k}_0=(k_0,0,0)$ for some $k_0\in\mathbb{R}$.

A single-mode coherent state with momentum $\mathbf{k}_0$ and frequency $\omega_0 = | \mathbf{k}_0 |$ can be viewed as the limiting case of a Gaussian wave packet with vanishing momentum uncertainty. To make this explicit, consider a multimode amplitude function of the form
\begin{equation}\label{alpha_G}
\alpha(\mathbf{k}) = \alpha_0 \sqrt{G_{(\delta k^1)^2}( k^1 - k_0) G_{(\delta k^2)^2}( k^2 ) G_{(\delta k^3)^2}( k^3 )},
\end{equation}
where $ \alpha_0$ is an overall amplitude prefactor, so that $| \alpha_0 |^2$ is the total photon number, and $\delta k^i$ denotes the uncertainty in the $i$-th momentum component. The ideal single-mode state is recovered in the limit $\delta k^i \to 0 $, which is practically realized when all other relevant physical scales are much larger than the momentum uncertainties. In this regime, the amplitude function becomes
\begin{equation}\label{alpha_delta}
\alpha(\mathbf{k}) = \alpha_0  \frac{\delta^3(\mathbf{k} - \mathbf{k}_0)}{\sqrt{V_\text{coh}}},
\end{equation}
where the coherence volume $V_\text{coh} = 1 / (8 \pi)^{3/2} \delta k^1  \delta k^2  \delta k^3 $ quantifies the scale over which the Fourier transform of $ \alpha(\mathbf{k})$ is distributed. Equivalently, in terms of the frequency uncertainty $\delta \omega$ and the angular spread $\delta \Omega$ in momentum space, the coherence volume can be written as $V_\text{coh} = 1 / (8 \pi)^{3/2} \omega_0^2 \, \delta \omega \, \delta \Omega $.

Under these assumptions, Eq.~\eqref{W_2_f_d_Lambda_eta_f_d_2} reduces to the following form
\begin{widetext}
\begin{align}\label{W_2_f_d_Lambda_eta_f_d_2_k}
W_2[\chi f_0,\chi f_0\circ \Lambda_1(\eta)]  = & \frac{1}{(2\pi)^6 \omega_0 V_\text{coh}} \int_{\mathbb{R}^3}  \frac{d^3 \mathbf{k}}{\omega(\mathbf{k})} \int_{\mathbb{R}^3} d^3 \mathbf{x} \chi(0, \mathbf{x}) \left[ \omega_0 \Im \!\left(\alpha_0 e^{i k_0 x^1}\right)  + i \omega(\mathbf{k}) \Re \!\left(\alpha_0 e^{i k_0 x^1}\right)  \right] e^{i \mathbf{k} \cdot \mathbf{x}} \nonumber \\
& \times  \int_{\mathbb{R}^3} d^3 \mathbf{y} \chi(0, \mathbf{y}) \left\lbrace \omega_0 \Im \!\left(\alpha_0 e^{i k_0 y^1}\right)  - i \left[ \cosh (\eta) \omega(\mathbf{k}) + \sinh (\eta) k^1 \right] \Re \!\left(\alpha_0 e^{i k_0 y^1}\right)  \right\rbrace\nonumber \\
& \times   \exp \! \left\lbrace -i \left[ \sinh (\eta) \omega(\mathbf{k}) + \cosh (\eta) k^1 \right] y^1 - i\mathbf{k}_\perp \cdot \mathbf{y}_\perp \right\rbrace.
\end{align}
\end{widetext}

The bump function $\chi$ appearing in Eq.~\eqref{W_2_f_d_Lambda_eta_f_d_2_k} remains to be specified. A concrete realization can be constructed by using the standard smooth transition function
\begin{equation}\label{vartheta}
\vartheta(s) = \frac{\Phi(s)}{\Phi(s) + \Phi(1 - s)},
\end{equation}
where
\begin{equation}
\Phi(s) = \begin{cases}
\exp \!\left( - \frac{1}{s} \right), & \text{ if } s>0, \\
0 , & \text{ if } s \leq 0.
\end{cases}
\end{equation}
This function $\vartheta(s)$ gives a smooth transition from $0$ to $1$ within the interval $0 \leq s \leq 1$. Based on this, we define a spatial bump function $\chi$ at time $x^0 = 0$ by
\begin{align}\label{chi_l_L_lambda_Lambda}
& \chi(0, \mathbf{x}) = \vartheta \!\left(1 +  \frac{l + \tau}{\delta l} - \frac{2}{\delta l} \left| x^1 + \frac{l + \tau + \delta l}{2} \right| \right) \nonumber \\
& \times \vartheta \!\left(1 + \frac{L + \tau}{\delta L} - \frac{2 \left| x^2\right|}{ \delta L} \right)  \vartheta \!\left(1 + \frac{L + \tau}{\delta L} - \frac{2 \left| x^3 \right|}{ \delta L} \right).
\end{align}
This function is smooth and compactly supported, matching the desired spatial profile of the enlarged detector region while ensuring a smooth transition to zero outside its support. To highlight its dependence on the geometric parameters and to avoid repeatedly specifying the initial-time condition $x^0 = 0$, we write $\chi(l + \tau, L + \tau, \delta l, \delta L; \mathbf{x})$ instead of $\chi(0, \mathbf{x})$. 

%\subsection{Rescaling with respect to $l + \tau$ and $L + \tau$}\label{Rescaling_with_respect_to_L}
%\subsection{Using dimensionless variables}\label{Rescaling_with_respect_to_L}
\subsection{Reformulation in dimensionless variables}\label{Rescaling_with_respect_to_L}

Let us introduce the dimensionless variables $\Delta \varphi$, $\tilde{\omega}_0$, $\widetilde{\delta l}$, $\widetilde{\delta L}$, $\tilde{\mathbf{x}}$, $\tilde{\mathbf{y}}$, $\tilde{\mathbf{k}}$, defined by rescaling the dimensionful quantities $k_0$, $\omega_0$, $\delta l$, $\delta L$, $\mathbf{x}$, $\mathbf{y}$, $\mathbf{k}$ with respect to $l + \tau$ or $L + \tau$ as follows: $\Delta \varphi = k_0 (l + \tau)$, $\tilde{\omega}_0 = \omega_0 (l + \tau)$, $\widetilde{\delta l} = \delta l/(l + \tau)$, $\widetilde{\delta L} = \delta L/(L + \tau)$, $\tilde{x}^1 = x^1/(l + \tau)$, $\tilde{\mathbf{x}}_\perp = \mathbf{x}_\perp/(L + \tau)$, $\tilde{y}^1 = y^1/(l + \tau)$, $\tilde{\mathbf{y}}_\perp = \mathbf{y}_\perp/(L + \tau)$, $\tilde{k}^1 = k^1 (l + \tau)$, $\tilde{\mathbf{k}}_\perp = \mathbf{k}_\perp (L + \tau)$. The first quantity, $\Delta \varphi$, represents the total optical phase accumulated by the wave as it traverses the region $\mathcal{V}_\text{det}$ of thickness $l + \tau$. That is, the ratio $\Delta \varphi / 2 \pi $ indicates how many wavelengths fit inside this region. The parameter $\tilde{\omega}_0$ is the mode frequency scaled by $l + \tau$. The third and fourth quantities, $\widetilde{\delta l}$ and $\widetilde{\delta L}$, describe the degree of smoothness of the bump $\chi$. The remaining variables, $\tilde{\mathbf{x}}$, $\tilde{\mathbf{y}}$, $\tilde{\mathbf{k}}$, serve as dummy integration variables in Eq.~\eqref{W_2_f_d_Lambda_eta_f_d_2_k}.

In addition to the dimensionless quantities discussed above, we introduce the parameter $N = |\alpha_0|^2 V_\text{det} / V_\text{coh}$, where $V_\text{det} = (l + \tau) (L + \tau)^2$ is the volume of $\mathcal{V}_\text{det}$ and $V_\text{coh} $ is the coherence volume characterizing the scale over which the test function $f_0$ is significantly distributed. Although $f_0$ has support across all space, one can adopt a practical, approximate notion of localization by identifying the region where its amplitude is non-negligible, effectively excluding the tails. In this sense, $V_\text{coh}$ quantifies how spread out the profile of the coherent state is. Accordingly, the ratio $ V_\text{det} / V_\text{coh}$ compares the size of the detection volume (set by $\tau$, $ l$ and $ L $) with the effective extent of the coherent state. Recalling that $|\alpha_0|^2$ represents the total photon number of the coherent state, the product $N = |\alpha_0|^2 V_\text{det} / V_\text{coh}$ can be interpreted as an estimate of the number of photons effectively ``seen'' by the detector\footnote{This interpretation is necessarily approximate: within AQFT there is no strict notion of particles localized in finite regions. Thus, viewing $N$ as the photon count within the detector relies on a looser, wavepacket-based notion of localization, the same that justifies interpreting $ V_\text{coh}$ as the effective volume occupied by the coherent state.}.

We also introduce the aspect ratio of $\mathcal{V}_\text{det}$ as $a = (l+\tau)/(L+\tau)$, which allows us to define the dimensionless frequency function as
\begin{equation}\label{omega_tilde}
\tilde{\omega}\!\left(\tilde{\mathbf{k}}\right) = \sqrt{\left|\tilde{k}^1\right|^2 + a^2 \left|\tilde{\mathbf{k}}_\perp\right|^2}.
\end{equation}
With these definitions, Eq.~\eqref{W_2_f_d_Lambda_eta_f_d_2_k} takes the form
\begin{widetext}
\begin{align}\label{W_2_f_d_Lambda_eta_f_d_2_k_adimensional}
& W_2[\chi f_0,\chi f_0\circ \Lambda_1(\eta)] \nonumber \\
= & \frac{N}{(2\pi)^6 \tilde{\omega}_0} \int_{\mathbb{R}^3}  \frac{d^3 \tilde{\mathbf{k}}}{ \tilde{\omega}(\tilde{\mathbf{k}})}  \int_{\mathbb{R}^3} d^3 \tilde{\mathbf{x}} \chi\!\left(1, 1, \widetilde{\delta l}, \widetilde{\delta L}; \tilde{\mathbf{x}}\right) \left\lbrace \tilde{\omega}_0 \Im \!\left[e^{i \arg (\alpha_0) + i \tilde{x}^1 \Delta \varphi }\right]  + i \tilde{\omega}(\tilde{\mathbf{k}}) \Re \!\left[e^{i \arg (\alpha_0) + i \tilde{x}^1 \Delta \varphi}\right]  e^{i \tilde{\mathbf{k}} \cdot \tilde{\mathbf{x}}}  \right\rbrace\nonumber \\
& \times  \int_{\mathbb{R}^3} d^3 \tilde{\mathbf{y}} \chi\!\left(1, 1, \widetilde{\delta l}, \widetilde{\delta L}; \tilde{\mathbf{y}}\right) \left\lbrace \tilde{\omega}_0 \Im \!\left[ e^{i \arg (\alpha_0) + i \tilde{y}^1 \Delta \varphi}\right] - i \left[  \cosh (\eta) \tilde{\omega}(\tilde{\mathbf{k}}) + \sinh (\eta) \tilde{k}^1 \right] \Re \!\left[e^{i \arg (\alpha_0) + i \tilde{y}^1 \Delta \varphi}\right]  \right\rbrace\nonumber \\
& \times   \exp \! \left\lbrace -i \left[ \sinh (\eta) \tilde{\omega}(\tilde{\mathbf{k}}) + \cosh (\eta) \tilde{k}^1 \right] \tilde{y}^1 - i\tilde{\mathbf{k}}_\perp \cdot \tilde{\mathbf{y}}_\perp \right\rbrace.
\end{align}
\end{widetext}

To ensure that the bump function $\chi$ reflects the characteristic length scales of $\mathcal{O}_1$, we assume $\widetilde{\delta l} \approx \widetilde{\delta L} \approx 1$. As discussed in Sec.~\ref{Inequality_for_general_coherent_states}, the conditions $\widetilde{\delta l} \lesssim 1 $ and $ \widetilde{\delta L} \lesssim 1$ guarantee a sufficient degree of locality, ensuring that the Wightman function $W_2[\chi f_0,\chi f_0\circ \Lambda_1(\eta)]$ remains sensitive to the detector size. In the opposite scenario, where $\widetilde{\delta l} \gg 1 $ or $ \widetilde{\delta L} \gg 1$, the configuration becomes practically indistinguishable from a global measurement, making the upper limit in Eq.~\eqref{upper_bound_f_O_d_2} effectively useless. At the same time, the condition $\widetilde{\delta l} \gtrsim 1 $ and $ \widetilde{\delta L} \gtrsim 1$ are necessary to prevent divergences in the Wightman function, thereby ensuring that the bound in Eq.~\eqref{upper_bound_f_O_d_2} remains meaningful.

To see why the condition $\widetilde{\delta l} \ll 1$ or $ \widetilde{\delta L} \ll 1$ causes the Wightman function $W_2[\chi f_0,\chi f_0\circ \Lambda_1(\eta)]$ to diverge, observe that in this limit, the bump function $\chi$ effectively becomes a sharp characteristic function of the detector region. Since the Fourier transform of a rectangular window yields a sinc function, we find that the integral $\int_{\mathbb{R}^3} d^3 \tilde{\mathbf{x}} \chi(1, 1, \widetilde{\delta l}, \widetilde{\delta L}; \tilde{\mathbf{x}})  \exp (i \tilde{\mathbf{k}} \cdot \tilde{\mathbf{x}})$ asymptotically behaves like $\exp ( - i \tilde{k}^1 / 2) \operatorname{sinc}(\tilde{k}^1/2) \operatorname{sinc}(\tilde{k}^2/2) \operatorname{sinc}(\tilde{k}^3/2)$ for large $\tilde{\mathbf{k}}$. Hence, the integrand of Eq.~\eqref{W_2_f_d_Lambda_eta_f_d_2_k_adimensional} with respect to $\tilde{\mathbf{k}}$ behaves as $ (|\tilde{k}^1 |^2 + a^2 |\tilde{\mathbf{k}}_\perp |^2)^{1/2} / (\tilde{k}^1 \tilde{k}^2 \tilde{k}^3 )^2$ at high momentum. This leads to logarithmic divergences as any of the components $\tilde{k}^i$ tends to infinity, while the others are held fixed.

\subsection{Numerical results}\label{Numerical_results}

We compute the upper bound for the expectation value $P_\text{click} (f)$, as given by the right-hand side of Eq.~\eqref{upper_bound_f_O_d_2}, i.e.,
\begin{equation}\label{upper_bound_f_O_d_2_approx_B}
P_\text{click}^\text{(max)}(f) = \min_{\zeta>0} \left[ \mathcal{E}_\zeta (f) + \exp\! \left( \frac{\pi^2}{2\zeta}\right) \sqrt{P_\text{dark}} \right]^2
\end{equation}
for various values of the experimental parameters $N$, $\Delta \varphi$, $a$, $\arg(\alpha_0)$ and $P_\text{dark}$. Here, $N =  |\alpha_0|^2 V_\text{det} / V_\text{coh}$ estimates the number of photons effectively ``seen'' by the detector, with $V_\text{det} = (l + \tau) (L + \tau)^2$ the volume of the extended detection region $\mathcal{V}_\text{det}$ and $V_\text{coh} = 1 / (8 \pi)^{3/2} \omega_0^2 \, \delta \omega \, \delta \Omega $ the coherence volume. The parameter $\Delta \varphi = k_0 (l+\tau)$ is the total optical phase accumulated across the thickness of $\mathcal{V}_\text{det}$, $a = (l + \tau)/(L + \tau)$ is the aspect ratio of $\mathcal{V}_\text{det}$, $\arg(\alpha_0)$ is the phase of the coherent amplitude $\alpha_0$, and $P_\text{dark}$ is the dark-count probability, i.e., the vacuum expectation value of the measurement operator $\hat{E}_\text{click}$.

Following the discussion of Sec.~\ref{Rescaling_with_respect_to_L}, we assume that $\widetilde{\delta l} \approx \widetilde{\delta L} \approx 1$. Ideally, the parameters $\widetilde{\delta l}$ and $ \widetilde{\delta L} $ would be selected to minimize the bound $P_\text{click}^\text{(max)}(f)$. However, doing so would require optimizing the right-hand side of Eq.~\eqref{upper_bound_f_O_d_2_approx_B} over both $\zeta$ and the pair $\widetilde{\delta l}$, $\widetilde{\delta L}$, thereby increasing computational complexity. To simplify the analysis while remaining consistent with the assumptions, we fix $\widetilde{\delta l} = \widetilde{\delta L} = 1$. Note that this choice does not constitute an approximation of a well-defined physical quantity, but rather reflects the use of a suboptimal configuration in realizing the upper bound given in Eq.~\eqref{upper_bound_f_O_d_2}.

\begin{figure}[h]
\includegraphics[width=\columnwidth]{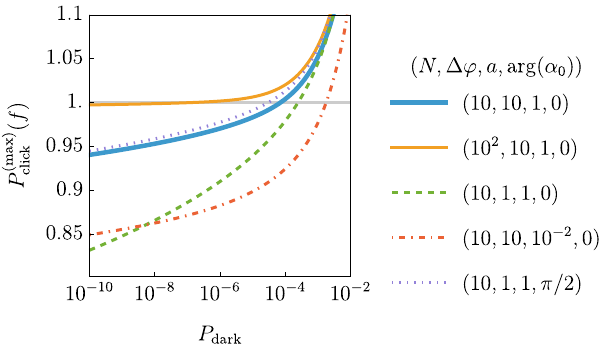}
\caption{Upper bound $P_\text{click}^\text{(max)}(f)$ on the click probability $P_\text{click}(f)$ as a function of the dark count $P_\text{dark}$. %The detector region is modeled as a square prism of dimensions $l \times L \times L$ operating over a finite time window $\tau$, while the state is a normally incident single-mode coherent state, with amplitude $\alpha_0$, momentum $k_0$ and coherence volume $V_\text{coh}$.
Results are shown for different combinations of parameters $(N, \Delta \varphi, a, \arg(\alpha_0))$, where $N  =  |\alpha_0|^2 (l + \tau) (L + \tau)^2 / V_\text{coh}$ estimates the number of photons effectively ``seen'' by the detector, $\Delta \varphi = k_0 (l+\tau)$ is the total optical phase accumulated across the thickness of the extended detection volume $\mathcal{V}_\text{det}$ of size $(l + \tau) \times (L + \tau) \times (L + \tau)$, $a = (l + \tau)/(L + \tau)$ is the aspect ratio of $\mathcal{V}_\text{det}$, $\arg(\alpha_0)$ is the phase of the coherent amplitude $\alpha_0$. The plots are displayed on a log--linear scale.
}\label{PlotBPclick}
\end{figure}

To compute the value of $P_\text{click}^\text{(max)}(f)$ for different combinations of the parameters $N$, $\Delta \varphi$, $a$, $\arg(\alpha_0)$ and $P_\text{dark}$, we numerically evaluate $\mathcal{E}_\zeta(f)$ using Eqs.~\eqref{epsilon_zeta_F_2} and \eqref{W_2_f_d_Lambda_eta_f_d_2_k_adimensional} and subsequently find the minimum in Eq.~\eqref{upper_bound_f_O_d_2_approx_B} for each configuration. The corresponding results are summarized in Fig.~\ref{PlotBPclick}. 

Following Ref.~\cite{falcone2025inequalityrelativisticlocalquantum}, the bound $P_\text{click}^\text{(max)}(f)$ should be compared with the ideal click probability, $P_\text{click}^\text{(ideal)} $, defined as the expectation value of $\hat{P}_\Omega^\perp = \hat{\mathbb{I}} - | \Omega  \rangle \langle\Omega |$. The operator $\hat{P}_\Omega^\perp$ projects onto the subspace orthogonal to the vacuum and thus represents an idealized detector: it has zero expectation value in the vacuum and equals $1$ for any state orthogonal to it. Since $\hat{P}_\Omega^\perp$ is not an element of any local operator algebra, it cannot correspond to a physically realizable measurement. The ratio $P_\text{click}^\text{(max)}(f)/P_\text{click}^\text{(ideal)} $ therefore provides a bound on how close one can come to the ideal click probability.

The ideal probability $P_\text{click}^\text{(ideal)}(f) = \langle f | \hat{P}_\Omega^\perp | f \rangle$ is explicitly given by $P_\text{click}^\text{(ideal)}(f) = 1 - \exp \! \left[ -W_2 (f,f) \right]$. The smeared two-point Wightman function appearing here can be written in terms of the amplitude $\alpha(\mathbf{k})$ as $W_2(f,f) = \int d^3 \mathbf{k} | \alpha(\mathbf{k}) |^2$. In the present case, using the Gaussian profile defined in Eq.~\eqref{alpha_G}, this integral evaluates to $W_2(f,f) = | \alpha_0 |^2$, resulting in $P_\text{click}^\text{(ideal)}(f) = 1 - \exp(-| \alpha_0 |^2)$. Since $| \alpha_0 |^2$ corresponds to the total photon number in the coherent state, $P_\text{click}^\text{(ideal)}(f)$ is expected to be very close to $1$ in realistic experimental regimes. Accordingly, one may approximate $P_\text{click}^\text{(max)}(f)/P_\text{click}^\text{(ideal)} \approx P_\text{click}^\text{(max)}(f)$ and interpret $P_\text{click}^\text{(max)}(f)$ as a bound on how close one can come to the ideal click probability.

It is important to note that a value of $P_\text{click}^\text{(max)}(f)$ equal to $1$ does not imply that the detector is ideal: this would only be possible if $\hat{E}_\text{click}$ were itself a global operator. Rather, such a value merely indicates that the computed bound is too loose to provide meaningful information about the nonideality of the detector. A bound $P_\text{click}^\text{(max)}(f)$ exceeding $1$ offers no useful constraint in the regime where $\hat{E}_\text{click}$ is local. %For this reason, in Fig.~\ref{PlotBPclick}, we restrict the plotted range of $P_\text{dark}$ for each configuration $(N, \Delta \varphi, a, \arg(\alpha_0))$ to the regime where $P_\text{click}^\text{(max)}(f)/P_\text{click}^\text{(ideal)} \ll 1$. interval in which the ratio $P_\text{click}^\text{(max)}(f)/P_\text{click}^\text{(ideal)} $ remains strictly below $1$.

From Fig.~\ref{PlotBPclick} we see that, for any fixed configuration $(N, \Delta \varphi, a, \arg(\alpha_0))$, the bound $P_\text{click}^\text{(max)}(f) $ decreases as $P_\text{dark}$ becomes smaller. This trend follows from Eq.~\eqref{upper_bound_f_O_d_2_approx_B}, where, in the limit of vanishing  $P_\text{dark}$, the optimal value of $\zeta$ shifts toward smaller values, resulting in a lower bound $P_\text{click}^\text{(max)}(f)$. This behavior clearly reflects the trade-off between quasi-ideal detector properties discussed in Ref.~\cite{falcone2025inequalityrelativisticlocalquantum}: suppressing false positives in the vacuum (lower $P_\text{dark}$) necessarily tightens the allowed response to excitations (lower $P_\text{click}^\text{(max)}(f)$). The idealized scenario in which the detector never clicks in the vacuum ($P_\text{dark} = 0$) yet still registers a definite response in the presence of a coherent state ($P_\text{click}^\text{(max)}(f) \neq 0$) is fundamentally unattainable.

Recall that $N = |\alpha_0|^2 V_\text{det} / V_\text{coh}$ provides an estimate of the number of photons effectively ``seen'' by the detector. By comparing the profile of $P_\text{click}^\text{(max)}(f)$ for $(N, \Delta \varphi, a, \arg(\alpha_0)) = (10,10,1,0)$ and $(10^2,10,1,0)$ in Fig.~\ref{PlotBPclick}, we find that increasing $N$ while keeping $\Delta \varphi$, $a$, $\arg(\alpha_0)$ and $P_\text{dark}$ fixed raises the profile of $P_\text{click}^\text{(max)}(f)$. Physically, this is consistent with interpreting $N$ as an approximate photon number within the detector region: as $N$ grows, the detection volume $V_\text{det}$ becomes large relative to the inverse effective photon density $V_\text{coh} / |\alpha_0|^2$, so the detector appears increasingly indistinguishable from an ideal, infinitely extended one from the perspective of the coherent state.

By comparing the configurations $(N, \Delta \varphi, a, \arg(\alpha_0)) = (10,10,1,0)$ and $(10,1,1,0)$ we observe that decreasing $\Delta \varphi$ lowers the bound $P_\text{click}^\text{(max)}(f)$. This is consistent with viewing $\Delta \varphi / 2 \pi $ as the number of wavelengths accommodated across the thickness of $\mathcal{V}_\text{det}$: when $\Delta \varphi$ is small, only a few wavelengths fit inside $\mathcal{V}_\text{det}$, rendering the coherent state effectively undetectable. Since $\mathcal{V}_\text{det}$ depends both on the detector dimensions (set by $l$ and $L$) and on the detector's operating time $\tau$ through $V_\text{det} = (l + \tau) (L + \tau)^2$, the result can be interpreted as follows: a small $\Delta \varphi$ corresponds to a situation where the detector spans too few wavelengths in thickness and, moreover, the measurement time is too short for the detector to ``collect'' enough wave cycles to register the field.

To examine how $P_\text{click}^\text{(max)}(f) $ depends on the geometrical features of the detection region, we study its behavior as the aspect ratio $a = (l+\tau)/(L+\tau)$ is varied while the remaining parameters are held fixed. Figure~\ref{PlotBPclick} shows the representative cases $(N, \Delta \varphi, a, \arg(\alpha_0)) = (10,10,1,0)$ and $ (10,10,10^{-2},0)$. We find that for small values of $a$, the profile of $P_\text{click}^\text{(max)}(f) $ decreases. This suggests that thin detection volumes with $l \ll L$ and relatively short operation times $\tau \ll L$ behave less ideally.

Lastly, the configurations $(N, \Delta \varphi, a, \arg(\alpha_0)) = (10,10,1,\pi/2)$, not shown in the plot, essentially overlaps with the curve for $(N, \Delta \varphi, a, \arg(\alpha_0)) = (10,10,1,0)$. This suggests that, for sufficiently large $\Delta \varphi$ the results are largely insensitive to the phase $\arg(\alpha_0)$. By contrast, when $\Delta \varphi \lesssim 1$, a clear difference emerges between $(N, \Delta \varphi, a, \arg(\alpha_0)) = (10,1,1,0)$ and $ (10,1,1,\pi/2)$. This is consistent with intuition: if the coherent state accumulates many phase cycles (large $\Delta \varphi$), the initial phase is effectively washed out, whereas for small phase accumulation ($\Delta \varphi \lesssim 1$), the outcome depends on the phase at which the coherent state enters.

\section{Conclusions}\label{Conclusions}

In this work, we examined the experimental prospects of a relativistic inequality that limits the performance of local detectors by enforcing a trade-off between vacuum insensitivity and responsiveness to excitations. Building on the Reeh--Schlieder approximation for coherent states, we derived an explicit, practically usable version of the bound for arbitrary coherent states. We then reformulated it in terms of the standard multimode amplitude $\alpha(\mathbf{k})$, aligning the result with standard quantum-optical language. Lastly, we illustrated the framework in a specific setup: a hyperrectangular detector acting over a finite time window and a normally incident single-mode coherent state, intended to approximate realistic photodetection and laser-beam settings.

Our results show that the bound retains the qualitative structure found in earlier idealized models \cite{falcone2025inequalityrelativisticlocalquantum}, while making the dependence on experimentally meaningful parameters explicit. The numerical study reveals a consistent pattern: reducing the dark-count probability inevitably tightens the achievable click probability, reflecting the fundamental impossibility of a perfectly vacuum-silent yet reliably excitation-sensitive local detector. The bound increases with the effective detected photon number $N$, decreases when the accumulated optical phase $\Delta \varphi$ is reduced, or when the detector is thin and operated for short times, i.e., for small aspect ratio $a$. By contrast, it shows only weak dependence on the phase $\arg(\alpha_0)$ when $\Delta \varphi$ is sufficiently large. Overall, these observations suggest that meaningful experimental tests of the inequality are most promising in regimes where the detector size and operation time are small compared to the relevant coherence scales.

\section*{Acknowledgment}

We acknowledge support from HORIZON EIC-2022-PATHFINDERCHALLENGES-01 HEISINGBERG Project No.~101114978.

\bibliography{bibliography}

\end{document}